\documentclass[a4paper,11pt]{article}
\pdfoutput=1 % if your are submitting a pdflatex (i.e. if you have
\usepackage{jheppub} % for details on the use of the package, please
\usepackage[T1]{fontenc} % if needed

\def\cs{$^{137}$Cs~}

\def\pbten{$^{210}$Pb~}

\def\pota{$^{40}$K~}

\title{\boldmath Limits on Interactions between Weakly Interacting Massive Particles and Nucleons Obtained with NaI(Tl) Crystal Detectors}

\collaboration{KIMS Collaboration}

\author[a]{K.W.~Kim}
\author[b]{G. Adhikari}
\author[b]{P. Adhikari}
\author[c]{S.~Choi}
\author[a]{C.~Ha}
\author[d]{I.S.~Hahn}
\author[a]{E.J.~Jeon}
\author[c]{H.W.~Joo}
\author[a]{W.G.~Kang}
\author[e]{H.J.~Kim}
\author[a]{N.Y.~Kim}
\author[c]{S.K.~Kim}
\author[a,b]{Y.D.~Kim}
\author[a,f]{Y.H.~Kim}
\author[a]{Y.J.~Ko}
\author[a,1]{H.S.~Lee\note{Corresponding author}}
\author[a]{J.S.~Lee}
\author[e]{J.Y.~Lee}
\author[a]{M.H.~Lee}
\author[a]{D.S.~Leonard}
\author[a]{S.L.~Olsen}
\author[a,g]{B.J.~Park}
\author[h]{H.K.~Park}
\author[f]{H.S.~Park}
\author[a]{K.S.~Park}

\affiliation[a]{Center for Underground Physics, Institute for Basic Science (IBS), \\Daejon 34126, Korea}
\affiliation[b]{Department of Physics, Sejong University, \\Seoul 05006, Korea}
\affiliation[c]{Department of Physics and Astronomy, Seoul National University, \\Seoul 08826, Korea}
\affiliation[d]{Department of Science Education, Ewha Womans University, \\Seoul 03760, Korea} 
\affiliation[e]{Department of Physics, Kyungpook National University, \\Daegu 41566, Korea}
\affiliation[f]{Korea Research Institute of Standards and Science, \\Daejon 34113, Korea}
\affiliation[g]{University of Science and Technology (UST), \\Daejon 34113, Korea}
\affiliation[h]{Department of Accelerator Science, Korea University, \\Sejong 30019, Korea}

\emailAdd{kwkim@ibs.re.kr}
\emailAdd{adhikari.astro@gmail.com}
\emailAdd{pushpaparticle@gmail.com}
\emailAdd{choi@phya.snu.ac.kr}
\emailAdd{cha@ibs.re.kr}
\emailAdd{ishahn@ewha.ac.kr}
\emailAdd{ejjeon@ibs.re.kr}
\emailAdd{hwjoo1240@gmail.com}
\emailAdd{wgkang@ibs.re.kr}
\emailAdd{hongjoo@knu.ac.kr}
\emailAdd{nykim@ibs.re.kr}
\emailAdd{skkim@snu.ac.kr}
\emailAdd{ydkim@ibs.re.kr}
\emailAdd{yhk@ibs.re.kr}
\emailAdd{yjko@ibs.re.kr}
\emailAdd{hyunsulee@ibs.re.kr}
\emailAdd{jsahnlee@ibs.re.kr}
\emailAdd{jylee8875@gmail.com}
\emailAdd{mhlee@ibs.re.kr}
\emailAdd{dleonard@ibs.re.kr}
\emailAdd{solsensnu@gmail.com}
\emailAdd{pbj7363@gmail.com}
\emailAdd{hyangkyu@korea.ac.kr}
\emailAdd{hyeonseo@kriss.re.kr}
\emailAdd{heppark@ibs.re.kr}

\abstract{
Limits on the cross section for weakly interacting massive particles (WIMPs) elastic scattering on nuclei
in NaI(Tl) detectors at the Yangyang Underground Laboratory are obtained from a 2967.4~kg$\cdot$day
data exposure. The nuclei recoiling from the scattering process are identified by the pulse shape of
the scintillation light signals that they produce.  The data are consistent with a no nuclear-recoil
hypothesis, and WIMP-mass-dependent 90\% confidence-level upper-limits are set on WIMP-nuclei elastic
scattering cross sections. These limits partially exclude the DAMA/LIBRA allowed region for WIMP-sodium
interactions with the same NaI(Tl) target material. The 90\% confidence level upper limit on the
WIMP-nucleon spin-independent cross section is 3.26$\times$10$^{-4}$ pb for a WIMP mass of 10~GeV/c$^2$.
}

\begin{document} 
\maketitle
\flushbottom

\section{Introduction}
\label{sec:intro}
A number of astrophysical observations provide evidence that the dominant matter component
of the universe is not ordinary matter, but rather non-baryonic dark matter~\cite{Clowe:2006eq,Ade:2015xua}.
Theoretically favored dark matter candidates are weakly interacting massive
particles~(WIMPs)~\cite{PhysRevLett.39.165}, which were well motivated by supersymmetric
models~\cite{Jungman:1995df}.
Many direct searches for WIMP dark matter in deep underground laboratories have been performed and
failed to find evidence for a signal~\cite{bertone05}, with the notable exception of the DAMA experiment
(including DAMA/NaI and DAMA/LIBRA) that has been operating for more than 20 years and has consistently
reported a positive signal of annual modulation of event rates in Na(Tl)
crystals~\cite{Bernabei:2000,Bernabei:2008,Bernabei:2010,Bernabei:2013xsa}, with a statistical significance
of more than 9$\sigma$~\cite{Bernabei:2013xsa}. This significance has recently been strengthened with results
from an additional six-year data exposure~\cite{Bernabei:2018phase2}.
The DAMA signal has been interpreted as being due to WIMP-nucleon scattering in the context of the standard model
for Milky Way galaxy's WIMP dark matter halo~\cite{Savage:2008er}.
However, the inferred WIMP-nucleon cross sections in this case are in conflict with null observations
from experiments that use different target materials~\cite{sckim12,PhysRevLett.118.021303,Tan:2016zwf,Aprile:2017iyp}.
Considering the high significance of the DAMA signal as well as the broad impact of a positive observation of
WIMP dark matter, an independent verification of the DAMA signal using the same NaI(Tl) target material is required.
A number of efforts with NaI target materials aimed at achieving this goal are in
progress~\cite{deSouza:2016fxg,amare14A,sabre,adhikari16,Fushimi:2015sew,Angloher:2017sft,Adhikari:2017esn}.

In the event that these ongoing efforts verify the annual modulation of event rates in the NaI(Tl) crystals,
it will remain of particular interest to study the characteristics of the events that produce this modulation.
Although the recoil signals that the DAMA experiment are usually attributed to nuclear recoils (NR), the experiment
does not distinguish them from electron recoils (ER) and, thus, the results allow for interpretations by models
that attribute the modulation signals to WIMP-electron interactions~\cite{PhysRevD.90.035027,PhysRevD.93.115037}.
If the DAMA modulation signal is confirmed, experimental studies that can distinguish electron recoils from nuclear
recoils will be necessary.

The Korea Invisible Mass Search~(KIMS) collaboration has developed low-background NaI(Tl) crystals for the purpose
of reproducing the DAMA's modulation results~\cite{kwkim15,adhikari16}.  A good understanding of the residual
radioactive background contaminants of the crystals has been achieved~\cite{Adhikari:2017gbj}. The characteristics
of NR signals in NaI(Tl) have been measured with a monoenergetic neutron source and found to have time-profile
characteristics that are different than those from ER that are produced by $\gamma$-rays from a
\cs source~\cite{Lee:2015iaa}. Similar time-profile differences in CsI(Tl) crystal scintillation-light signals have
been used to discriminate between NR and ER events, so-called pulse shape discrimination (PSD), on a statistical basis
as described in Ref.~\cite{hslee07,sckim12}, and set upper limits on WIMP-nucleon cross sections in the context of the
standard galactic halo model~\cite{Lewin:1995rx}.  This paper presents results from a search for WIMP-nucleon
interactions with two, low-background NaI(Tl) crystals using a 2967.4~kg$\cdot$day exposure that uses a PSD analysis
to directly extract NR-induced events.

\section{Experiment}
The two crystals are cylindrical in shape but with different dimensions; one is 12.7~cm in diameter and 17.8~cm in
length with a mass of 8.26~kg (NaI1); the other is 10.7~cm in diameter and 27.9~cm in length with a mass of 9.15~kg
(NaI2), as described in Table~\ref{tab_crystal}. The crystals are encapsulated in a cylindrical copper structure with
quartz windows at each end. Scintillation-light signals are read out by 76~mm Hamamatsu R12669 photomultiplier tubes~(PMTs)
that are optically coupled to each end of the cylinder.  The assembled detector modules are installed in the
shielding structure that was used for the KIMS CsI(Tl) detector measurements at the Yangyang Underground
Laboratory (Y2L)~\cite{Lee:2005qr,kims_crys1,sckim12}.  The shield includes a 12-module array of low-background
CsI(Tl) detectors situated inside of a shield that is comprised, from inside out, of 10~cm of copper, 5~cm of
polyethylene, 15~cm of lead, and 30~cm of mineral oil to attenuate external radiation. Nitrogen gas was flushed
through the detector setup to avoid radon contamination and maintain stable temperature and humidity levels. The data
were collected between November~2013 and January~2015 with configurations described in Ref.~\cite{kwkim15}.

\begin{table}[tbp]
\centering
\begin{tabular}{|c|c|c|c|}
\hline
Name & Size (D)$\times$(L) & Mass  (kg) & Exposure (days)\\
\hline
NaI1 & 12.7 cm$\times$17.8 cm & 8.26 & 80.8 \\
NaI2 & 10.7 cm$\times$27.9 cm&  9.15 & 251.4\\
\hline
\end{tabular}
\caption{\label{tab_crystal}The crystals used in this analysis and the amount of data for each crystal}
\end{table}

Signals from each PMT are amplified by custom-made preamplifiers and digitized by a 400-MHz flash analog-to-digital
converter. An event is triggered when at least a single photoelectron is observed in each PMT within a 200~ns time
interval. The energy scales for the crystals are determined using 59.5~keV $\gamma$ events from a $^{241}$Am calibration
source.  The 68.7~keV line from cosmogenic $^{125}$I and the 46.5~keV line from a residual $^{210}$Pb contamination are
used to monitor the time-dependence of the energy scale.

The measured light yields were 15.5~photoelectrons/keV for both crystals, which is approximately twice as large
as that of the crystals used in previous studies by NAIAD~\cite{Alner:2005kt} and DAMA~\cite{BERNABEI1996757}. 
The increased light yield is the result of improvements in both the crystals and the PMTs. The crystals were grown with
a technique that was developed by Alpha Spectra Inc.~\footnote{http://www.alphaspectra.com} that results in improved light
yields.  In addition, high quantum efficiency, low-background PMTs developed by Hamamatsu
Photonics~\footnote{https://www.hamamatsu.com} were used.  These new PMTs, also used in DAMA/LIBRA-phase2, provide, on
average, about a 30\% improvement in light yield~\cite{bernabei12had}.  The high light yields of our detectors allow for
a discrimination of NR events using PSD methods~\cite{Lee:2015iaa}.

\begin{figure}[tbp]
\centering
\includegraphics[width=0.6\textwidth]{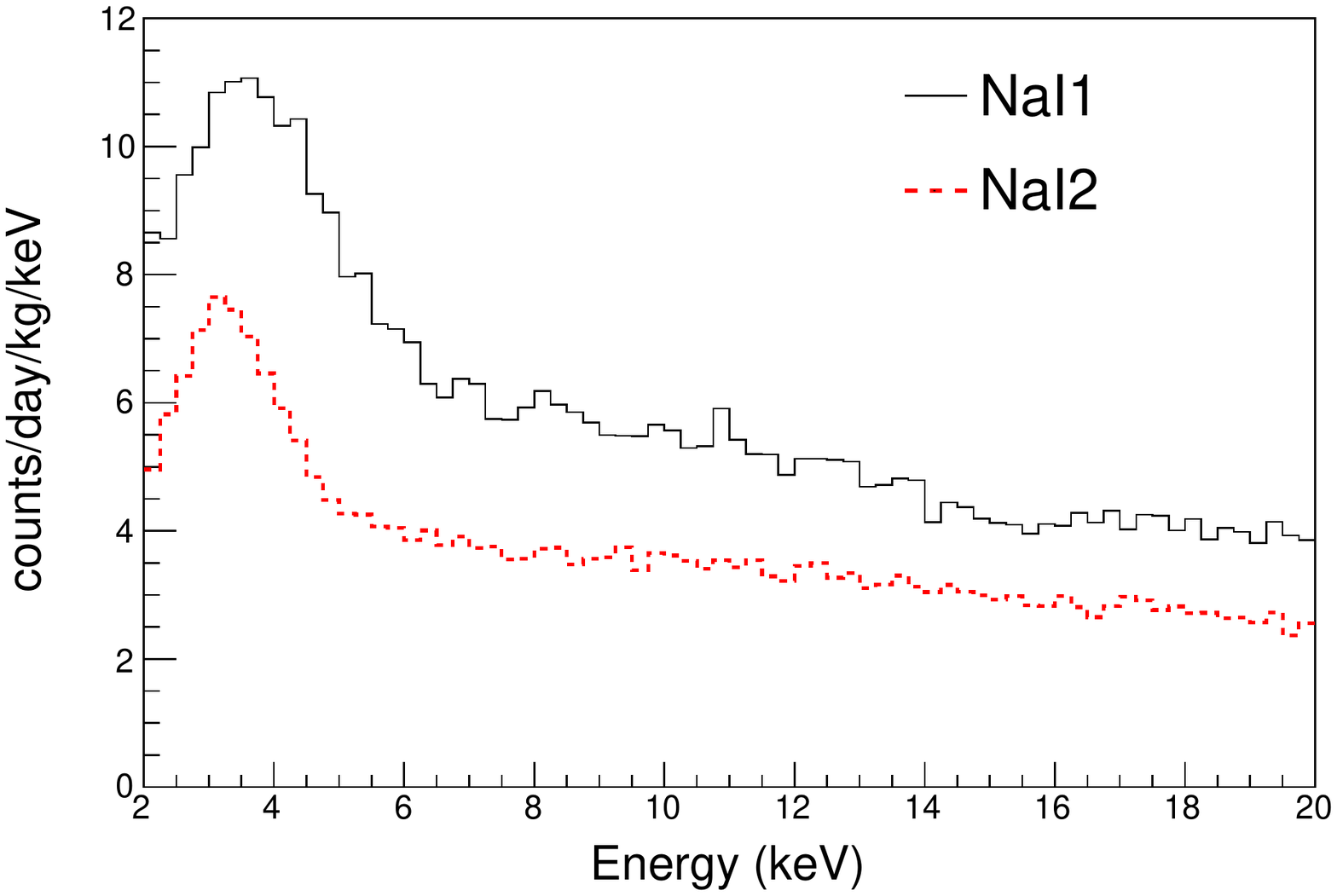}
\caption{\label{fig_E}
Energy spectra for NaI1 and NaI2 after the application of the event selection requirements described in the text. The main
backgrounds in the crystals are from \pota and \pbten contaminations, and cosmogenic activations as described in
Refs.~\cite{kwkim15,adhikari16}. 
}
\end{figure}

The background levels of the NaI1 (NaI2) crystals are approximately 9~(5)~counts/day/kg/keV in the 2--6~keV recoil
energy region and decrease to 6~(4)~counts/kg/day/keV in the 6--10~keV energy region, as shown in figure~\ref{fig_E}.
This figure shows the background spectra of the crystals after the application of the event selection criteria
described below.  These background levels are lower than those of the crystals used by the NAIAD
experiment~\cite{Alner:2005kt}, but higher than those in DAMA/LIBRA~\cite{Bernabei:2013xsa}. The higher background
in is primarily due to a larger contamination of \pbten~~\cite{kwkim15, adhikari16}. 
Detailed background studies of these crystals are reported elsewhere~\cite{kwkim15,adhikari16,Adhikari:2017gbj,cosinebg}.

In order to have good reference distributions for known backgrounds and signal events, we accumulated calibration data samples
for ER, NR, and surface $\alpha$ recoil~(SR) events. The ER event calibration sample was obtained with NaI1 and NaI2 by
irradiating them with a \cs source in the copper shielding chamber. The setup and running conditions for the ER calibration
sample were exactly same as the WIMP search data.

The characteristics of NR events were measured with a small NaI(Tl) crystal~(2$\times$2$\times$1.5~cm$^3$)~(referred to as NaI-NR)
that was taken from the same ingot as NaI2, by exposing it to a monoenergetic 2.43~MeV neutron beam from a deuteron-based
generator~\cite{qf_cosine}.  Neutrons that scattered from the NaI(Tl) crystal were detected in neutron tagging counters
containing BC501A liquid scintillator~\cite{hjkim04}.  A time coincidence between signals from the target crystal and any
one of the neutron detectors was used to select neutron scattering events. Neutron-induced signals in the tagging counters
were distinguished from electron and gamma background-induced events by their pulse shapes.  Details of neutron calibrations
with similar arrangements are reported in Ref.~\cite{Lee:2015iaa,qf_cosine,jhlee-qf15,hslee-psd14}.

It is known that $^{206}$Pb nuclei recoiling from $\alpha$ decays of $^{210}$Po atoms attached to the crystal surfaces
can mimic WIMP events in the recoil-energy region of interest~(ROI)~\cite{Kudryavtsev:2001hp,Kim:2011je,Strauss:2014hog}.
To isolate a calibration sample of SR events, we cut an 8~cm diameter, 10~cm long cylindrical NaI(Tl) crystal into two,
5~cm-long pieces and contaminated the exposed surface of one of them~(referred to as NaI-SR) with $^{222}$Rn. After waiting
two weeks to allow for the build-up of surface $^{210}$Po, we reconnected the contaminated and uncontaminated cylinders with
a thin light barrier, and tagged recoil $^{206}$Pb signals in NaI-SR by 5.3~MeV $^{210}$Po $\alpha$ signals in the
uncontaminated cylinder~\cite{Kim:2018kbs}.

\section{Analysis}
In order to discriminate PMT-induced noise from radiation-induced signal events, event selection criteria using the
separation parameters described in Ref.~\cite{kwkim15} are applied. The most important parameters reflect the differences
between the ``slow''~(100--600~ns) and ``fast''~(0--50~ns) charge sums and the charge asymmetry between the two PMTs.
A NR event from the WIMP interaction is expected to produce a hit in a single crystal, designated as a single-hit~(SH) event,
while events with accompanying hits in other surrounding NaI(Tl) or CsI(Tl) crystals are designated as multiple-hit~(MH) events.
The MH events are used to determine the selection efficiency and for modeling ER events.
The efficiency determined with the MH events is greater than 90\% in the ROI~(2--8~keV). A sample of NR events from the
neutron beam studies are used for an independent determination of the efficiency for NR events with
results that are consistent with the determination with the ER events to within 2\%.

We use the signal shapes to determine the contributions of WIMP-induced NR events.  To characterize the shape,
we first identify photon clusters as an isolated pulse that is typically equivalent to a single photoelectron
for signals in the low-energy region~\cite{Lee:2005qr}. With these clusters, we determine the natural logarithm of
the mean time (LMT) of an event waveform calculated over a 1~$\mu$s time window, from

\begin{equation}
                \log(\mathrm{Mean~Time}) = \mathrm{LMT} = \log\Big( \frac{\sum_{i=1}^n A_{i}t_{i}}{\sum_{i=1}^n A_{i}} -t_{1}\Big), 
\label{eq:meantime}
\end{equation}
where $A_{i}$ is the charge of the $i$th cluster and $t_i$ is the time of the $i$th cluster.

\begin{figure}[tbp]
\centering
\begin{tabular}{cc}
\includegraphics[width=0.45\textwidth]{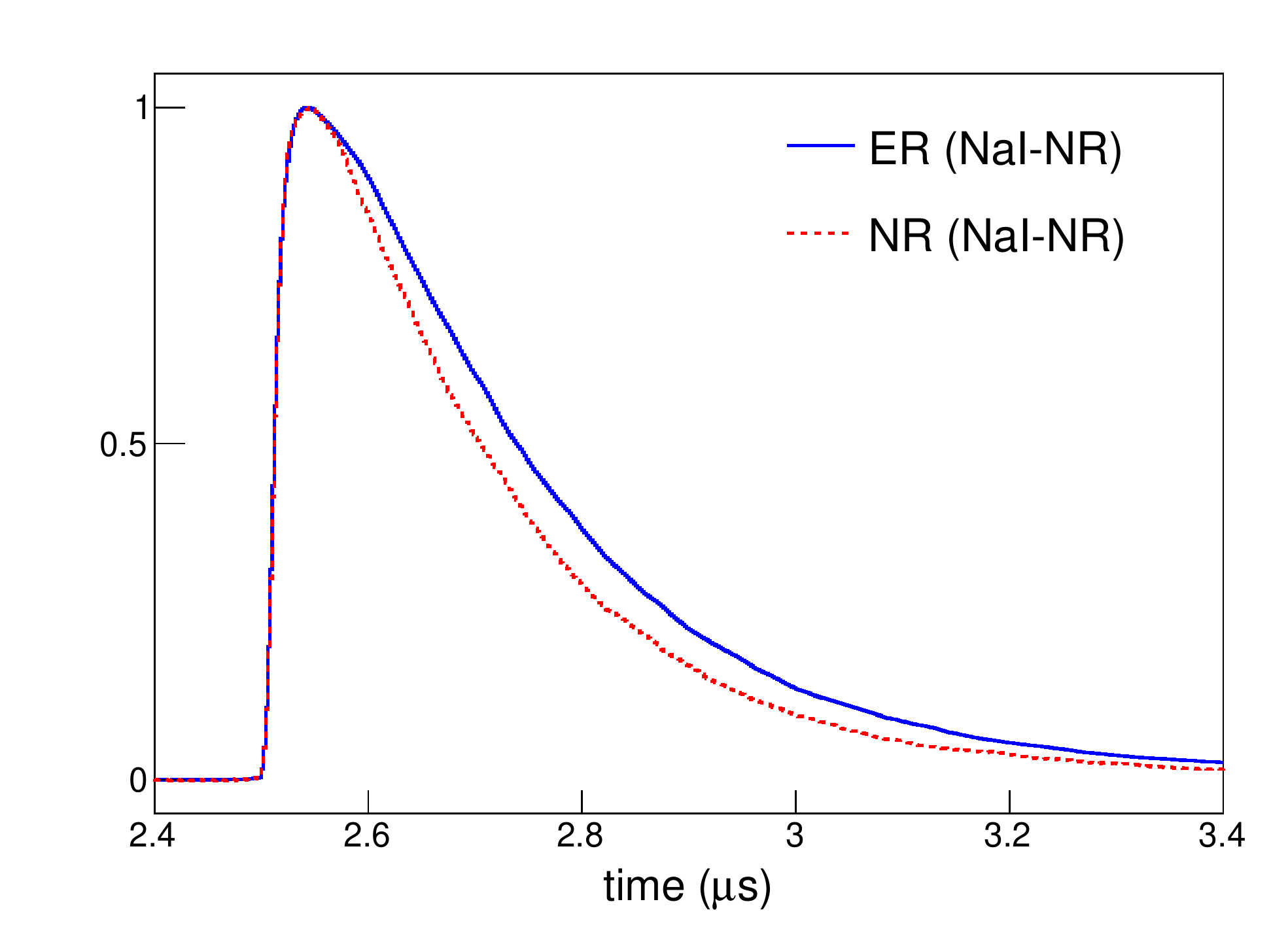} &
\includegraphics[width=0.45\textwidth]{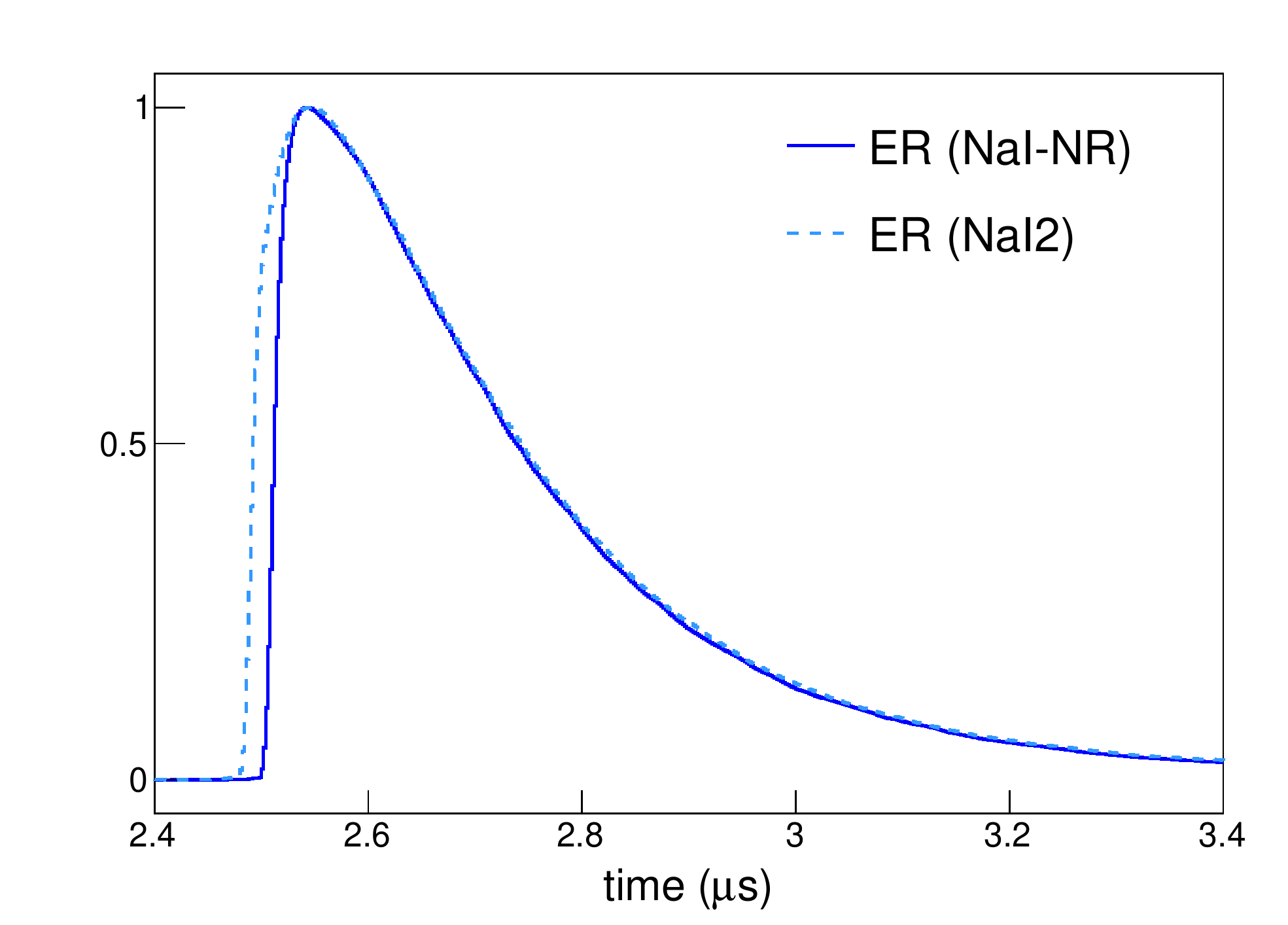} \\
(a) & (b) \\
\end{tabular}
\caption{\label{fig_wav}
  The averaged waveforms for NaI-NR and NaI2 signals in the 2 to 10 keV energy region.
  (a) The solid blue curve and dotted red curve represent averaged ER- and NR-induced waveforms in NaI-NR,
  respectively, where distinctly different decay times are evident.
  (b) The dashed sky-blue line is the averaged ER-induced waveform in NaI2, which shows a slower rise-time than
  that in NaI-NR (solid blue curve), but the same decay-time as that seen in the smaller crystal.
}
\end{figure}

Figure~\ref{fig_wav}(a) shows the averaged waveforms for ER- and NR-induced events with energies between 2 and 10 keV,
where differences in then decay portion of the signals are evident.  We can, therefore, use the LMT parameter to
characterize the two different types of recoils.  As summarized in Table~\ref{tab_lightyield}, the crystals used
to, produce the NR and SR calibration samples were different from those used for WIMP search measurements. 
Differences between the responses to ER-induced events for the WIMP-search crystal~(NaI2) and the small crystal used
to generate the  NR calibration sample~(NaI-NR) can be seen in figure~\ref{fig_wav}(b).  While the rise times in the
two crystals are different, the decay portion of the waveforms in the two different-sized crystals are well matched.
Light signals produced in larger crystals have longer light paths to the PMT and, thus, experience more photon absorption
and reemission, and undergo more reflections, resulting in longer rise times~\cite{knoll,weber:risetime}.
On the other hand, these effects are small in comparison to the scintillation light decay time.
The NR calibration was performed with small-sized crystal to minimize the effects of multiple scattering of neutrons
that would effect the measured LMT values.  Similar crystal-size effects on the LMT distribution are expected for the
SR calibration sample.

\begin{figure}[tbp]
\centering
\begin{tabular}{cc}
\includegraphics[width=0.45\textwidth]{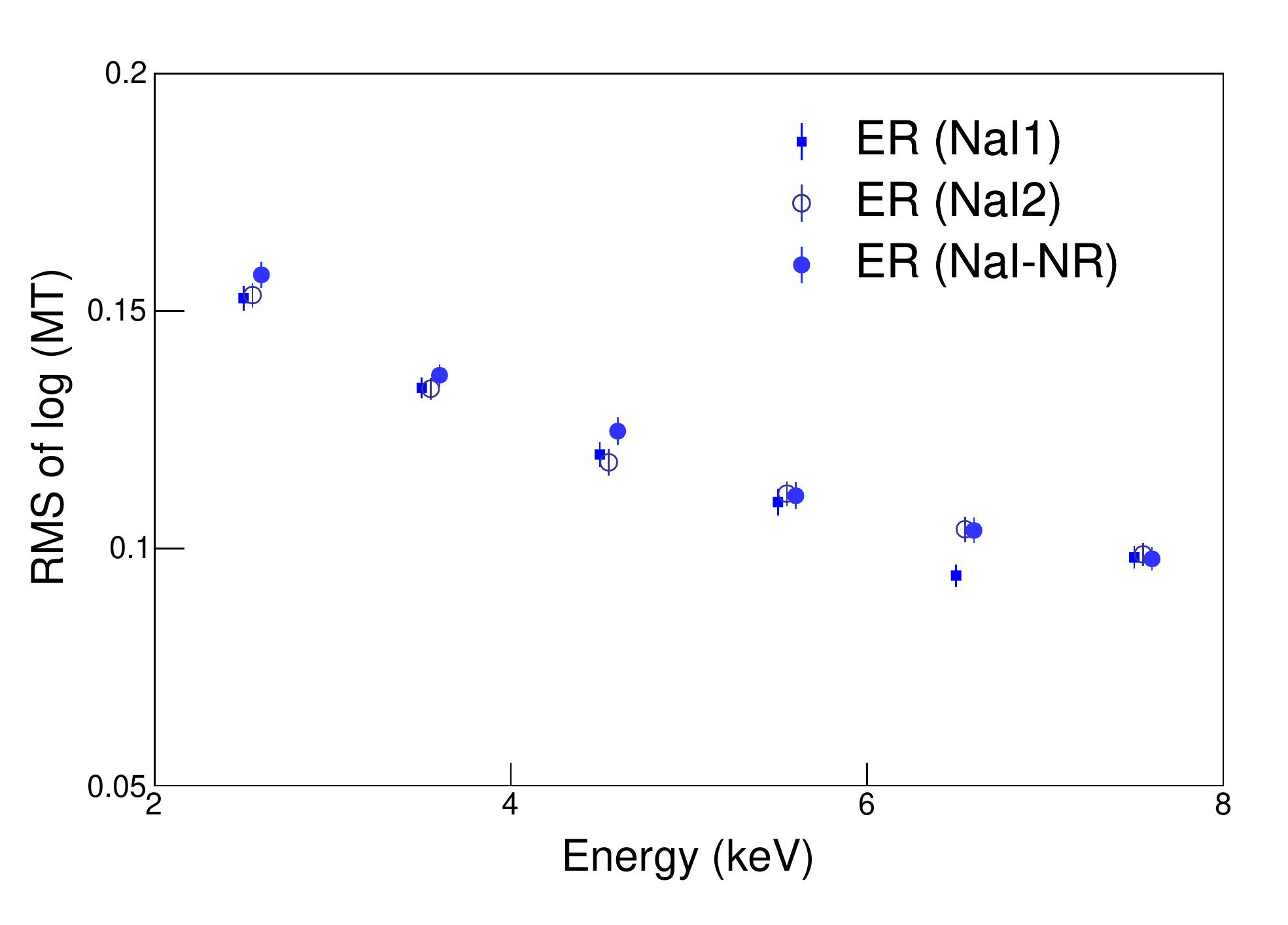} &
\includegraphics[width=0.45\textwidth]{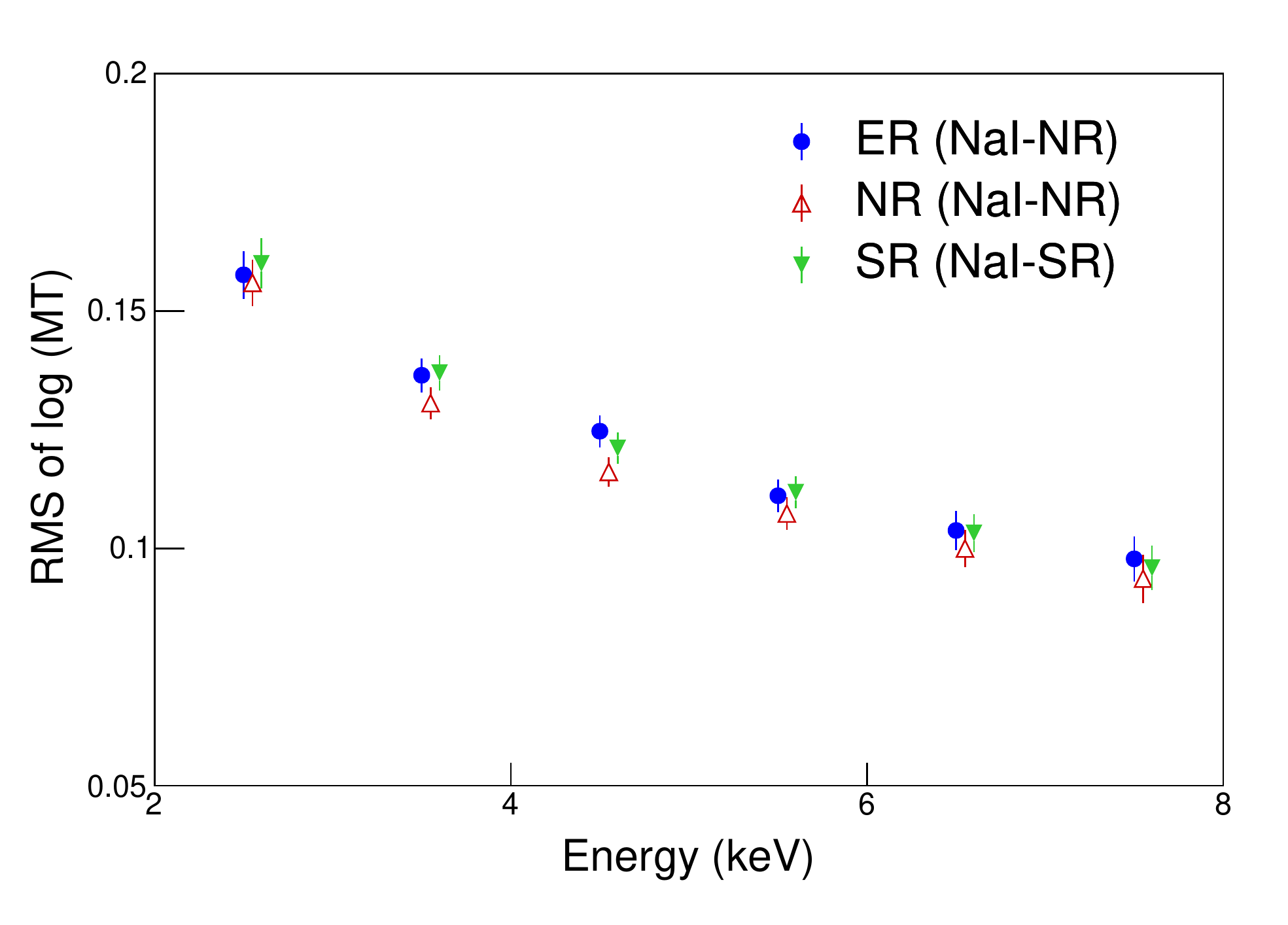} \\
(a) & (b)  \\
\end{tabular}
\caption{\label{fig_rms_mean}
  (a) The recoil energy dependence of the RMS values of LMT distributions for ER-induced signals in the crystals that were
  used for the underground WIMP-search and the above-ground calibration data samples. All of the ER samples have RMS values
  that are equal within the measurement errors, which indicates that the discrimination power is the same for the different
  crystals. 
  (b) The RMS values of the LMT distributions for the ER, NR and SR calibration samples.  Since the RMS depends on the crystal
  light yield, and the crystals used to produce the calibration samples have similar light yields, we expect each sample to
  have the same RMS at each energy. 
}
\end{figure}

The root mean square width~(RMS) of the LMT distributions determine the discrimination capability. 
Figure~\ref{fig_rms_mean} (a) shows that the RMS values of the LMT distributions for ER events from three different crystals
are mutually consistent.  Likewise,  Figure~\ref{fig_rms_mean}(b) shows that the RMS values for the ER, NR and SR calibration
samples are also mutually consistent.  The different, crystal-size-dependent rise times for the  NR and SR events are corrected
for by using  differences in the ER measurements in the corresponding crystals, assuming that the ratio $R_{\tau}=\tau_{n}/\tau_{e}$
is independent of crystal~\cite{ahmed03,Lee:2007af,sckim12}, where $\tau_{n}$ and $\tau_{e}$ are NR and ER decay times, respectively.

\begin{table}[tbp]
\centering
\begin{tabular}{|c|c|c|c|c|}
\hline
Name & LY (p.e./keV) & ingot & Mass (kg) & Usage\\
\hline
NaI1 & 15.6$\pm$1.4 & A & 8.26 &WIMP search \\
NaI2 & 15.5$\pm$1.4 & B & 9.15 &WIMP search\\
NaI-NR& 15.8$\pm$0.9 & B& 0.02 &NR calibration\\
NaI-SR& 14.4$\pm$1.6 & C& 0.64 &SR calibration\\
\hline
\end{tabular}
\caption{\label{tab_lightyield} The Light Yield (LY) and powder of crystals used in WIMP search data and calibrations}
\end{table}

%\begin{figure}[tbp]
%\begin{center}
%\begin{tabular}{ccc}
%\includegraphics[width=0.33\textwidth]{model_sr4.pdf} &
%\includegraphics[width=0.33\textwidth]{model_nr4.pdf} &
%\includegraphics[width=0.33\textwidth]{model_er4.pdf} \\
%(a) SR & (b) NR & (c) ER \\
%\end{tabular}
%\caption{
%The LMT distributions of 2--3~keV SR (a), NR (b) and ER (c) for NaI2 are presented. 
%Data~(points) are overlaid with modelings of the asymmetric gaussian function before~(dotted line) and after correction applied~(solid line). 
%}
%\label{fig_model}
%\end{center}
%\end{figure}

\begin{figure}[tbp]
\centering
\begin{tabular}{cc}
\includegraphics[width=0.45\textwidth]{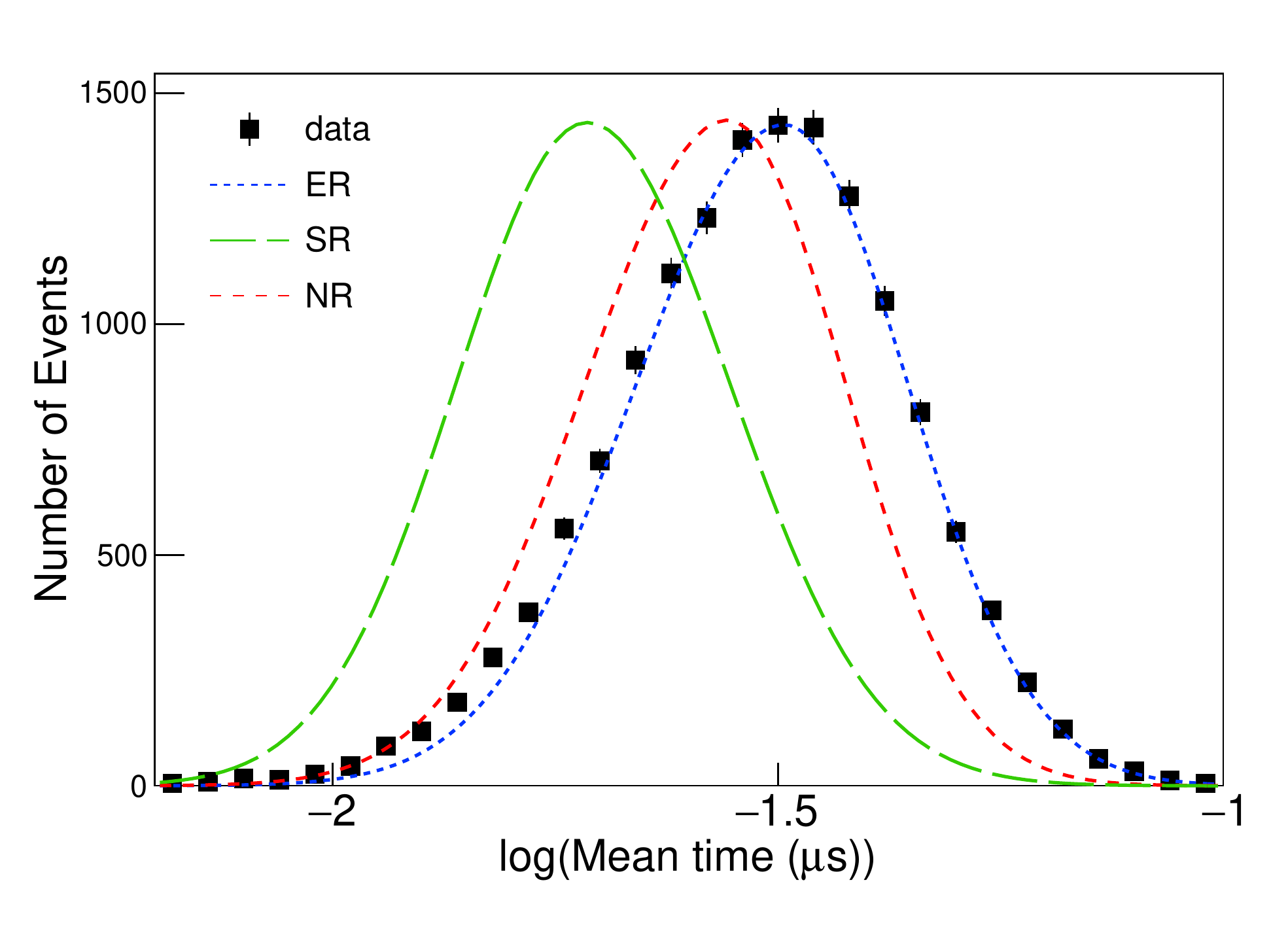} &
\includegraphics[width=0.45\textwidth]{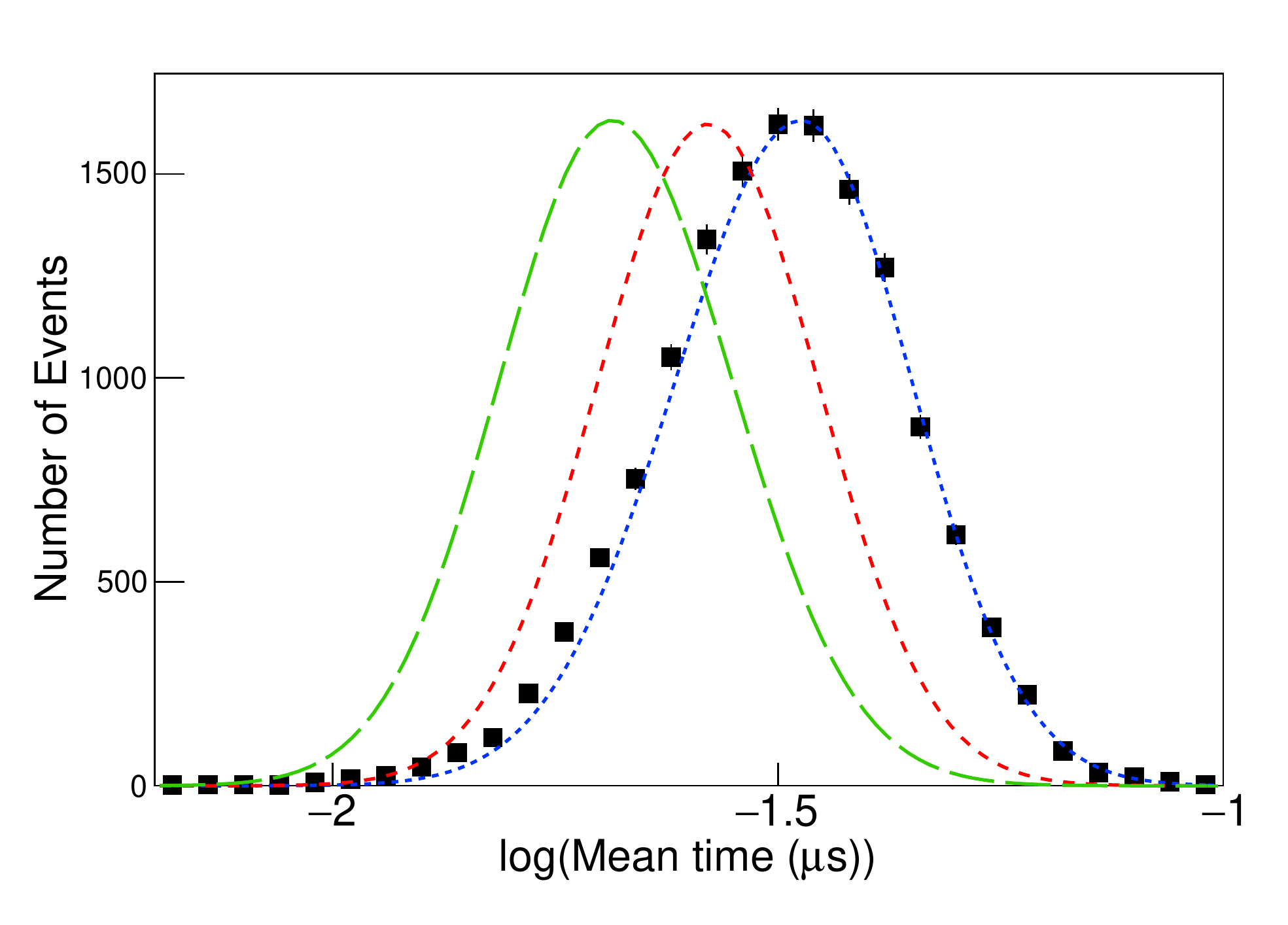} \\
(a) 2--3~keV & (b) 3--4~keV \\
\end{tabular}
\caption{\label{fig_mt}
The LMT distributions of data~(point) in NaI2 are compared with PDFs of ER~(dotted line), SR~(long dashed line), and NR~(dashed line) for events at 2--3~keV~(a) and 3--4~keV~(b).
}
\end{figure}

%Figure 1 shows a comparison of the LMT distributions for different recoil-type events, together with data in energies of 2{3 keV and 3{4 keV for NaI2. Models for the recoil events are built using an asymmetric Gaussian function for every 1 keV energy bin.
Corrected LMT distribution of each calibration data is modeled with an asymmetric gaussian function to obtain a probability density function~(PDF) for each 1~keV energy bin.% as one can see in figure~\ref{fig_model}. 
We compare the LMT distributions from WIMP search data in NaI2 with PDFs for each recoil type in figure~\ref{fig_mt}. As one can see in this plot, our data are dominantly described by ER, but they contain small amount of fast decaying components. 

\begin{figure}[tbp]
\centering
\begin{tabular}{cc}
\includegraphics[width=0.45\textwidth]{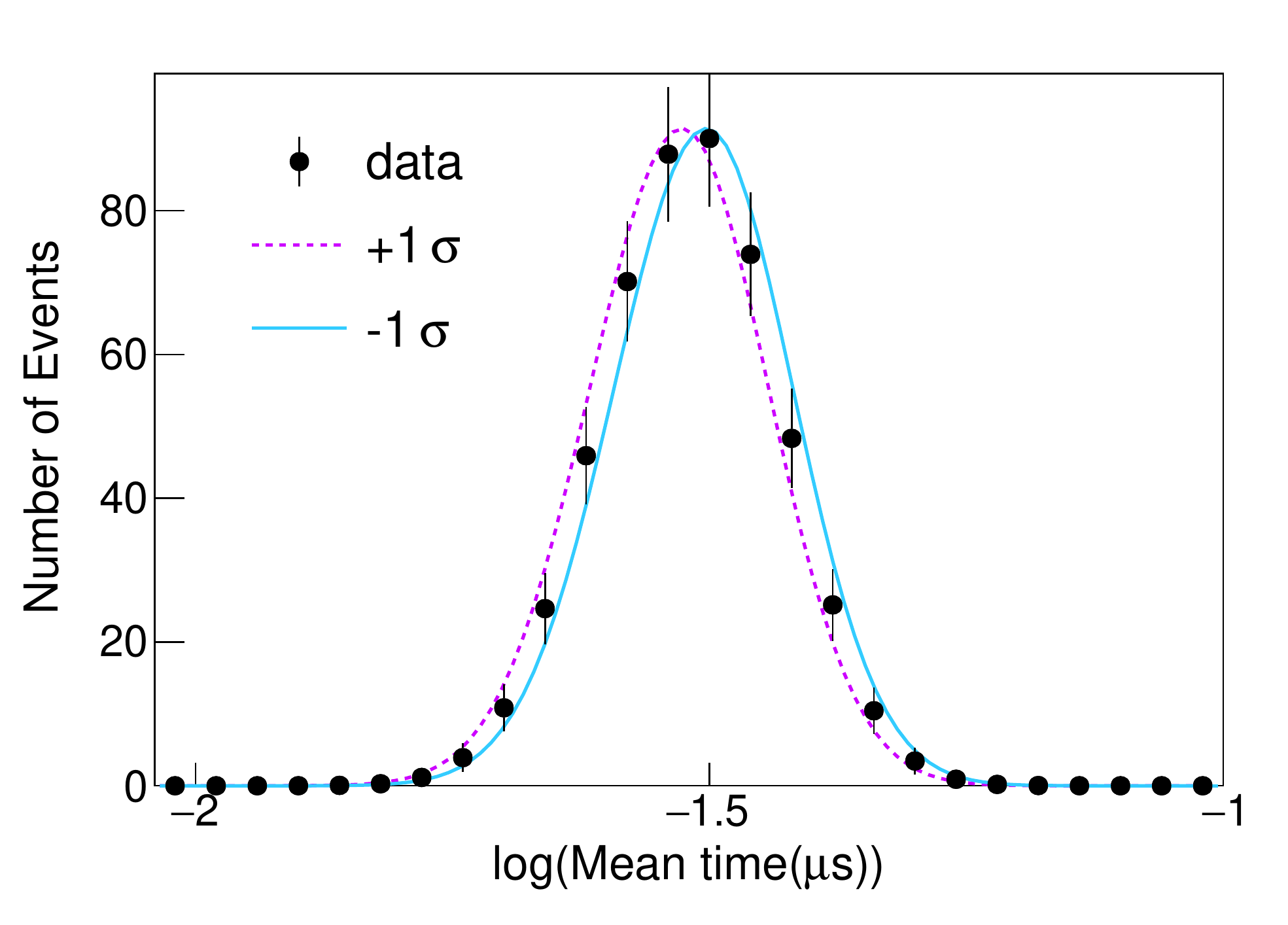} &
\includegraphics[width=0.45\textwidth]{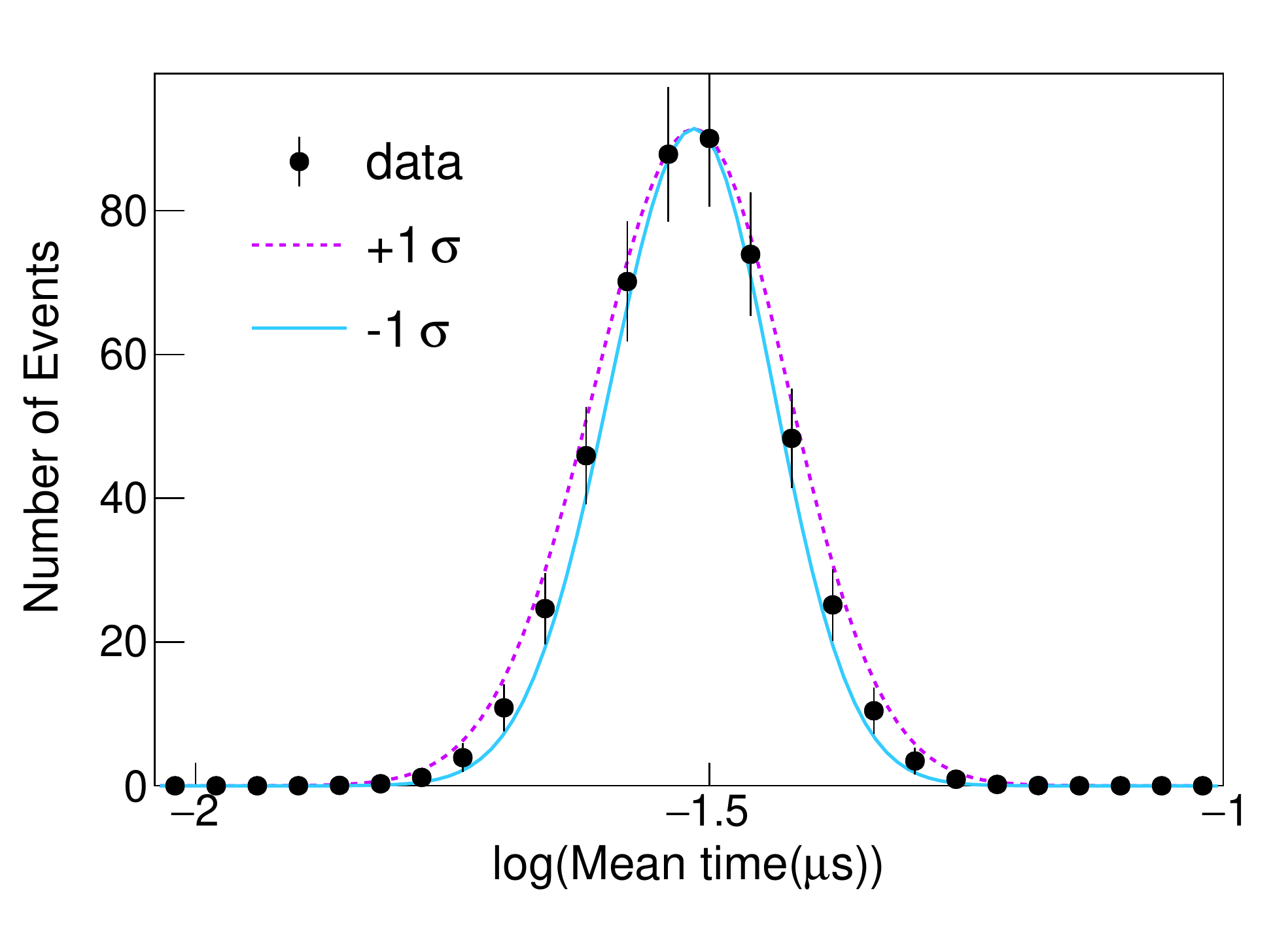} \\
(a) & (b)\\
\end{tabular}
\caption{\label{fig_syst}
The distributions shifted by the mean (a) and RMS (b) uncertainties for NR events in the 7 to 8 keV energy region, where the largest shape changes occur.
}
\end{figure}

Several sources of systematic uncertainties are considered in this analysis.
The dominant components are associated with mean and RMS uncertainties in the modeling of each recoil due to the limited statistics of the calibration data.
Allowed shape changes for PDF of the NR events are shown in figure~\ref{fig_syst}. 
PDF for each recoil type and each 1~keV bin has considered independently.
The uncertainty caused by corrections due to different distributions of the ER events caused by rise time differences from different size crystal measurements are also included. 
Uncertainties in the efficiency estimation and differences in the independent cross-check for different recoil-type samples are considered as possible rate changes.
Variations in the energy calibrations and resolutions are also included as systematic errors.

To search for WIMP-induced NR events in the data, binned maximum-likelihood fits to the measured LMT spectra between 2 and 8~keV are performed.
The Bayesian Analysis Toolkit~\cite{Caldwell:2008fw} is used with PDFs based on the LMT models for the NR, SR, and ER events shown in figure~\ref{fig_mt}.
The likelihood is built for each 1~keV bin for 2--8~keV of each crystal and multiplied as a single likelihood. Uniform priors are used for three different recoil types.
The numbers of NR events across different energy bins and crystals are constrained by using the shapes of simulated WIMP energy spectra.
The measured energy spectrum of the SR from Ref.~\cite{Kim:2018kbs} is used to constrain the energy-dependent rate of the SR events in each crystal. However, the total amount of SR events for each crystal is freely floated.

We obtain the expected energy spectrum for WIMP interactions by generating samples for 17 different WIMP masses, ranging from 5~GeV/c$^2$ to 1,000~GeV/c$^2$ using the same parameters that are used for the WIMP interpretation of the DAMA/LIBRA-phase1 result~\cite{Savage:2008er} in the context of the standard halo model~\cite{Lewin:1995rx}. We used the same quenching factor (the ratio of measured energies for nuclear and electron recoils of the same energy) measured by DAMA~\cite{damaqf} with a $^{252}$Cf neutron source of Q$_{\mathrm{Na}}$=0.3 and Q$_{\mathrm{Na}}$=0.09 even though recent measurements using mono-energetic neutron beams report significantly lower values~\cite{qf_collar,qf_xu,qf_cosine}. As noted in Ref.~\cite{qf_collar}, this is mainly caused by incorrect evaluation of old measurements so that it is reasonable to use the same factor for comparison of two experiments using the same NaI(Tl) target. 
The systematic uncertainties are included in the fit as nuisance parameters with Gaussian priors assumed for both the rate and the shape changes.
We consider the possibility of correlated rate and shape uncertainties as well as uncorrelated bin-by-bin statistical uncertainties.

\begin{figure}[tbp]
\centering
\begin{tabular}{cc}
\includegraphics[width=0.4\textwidth]{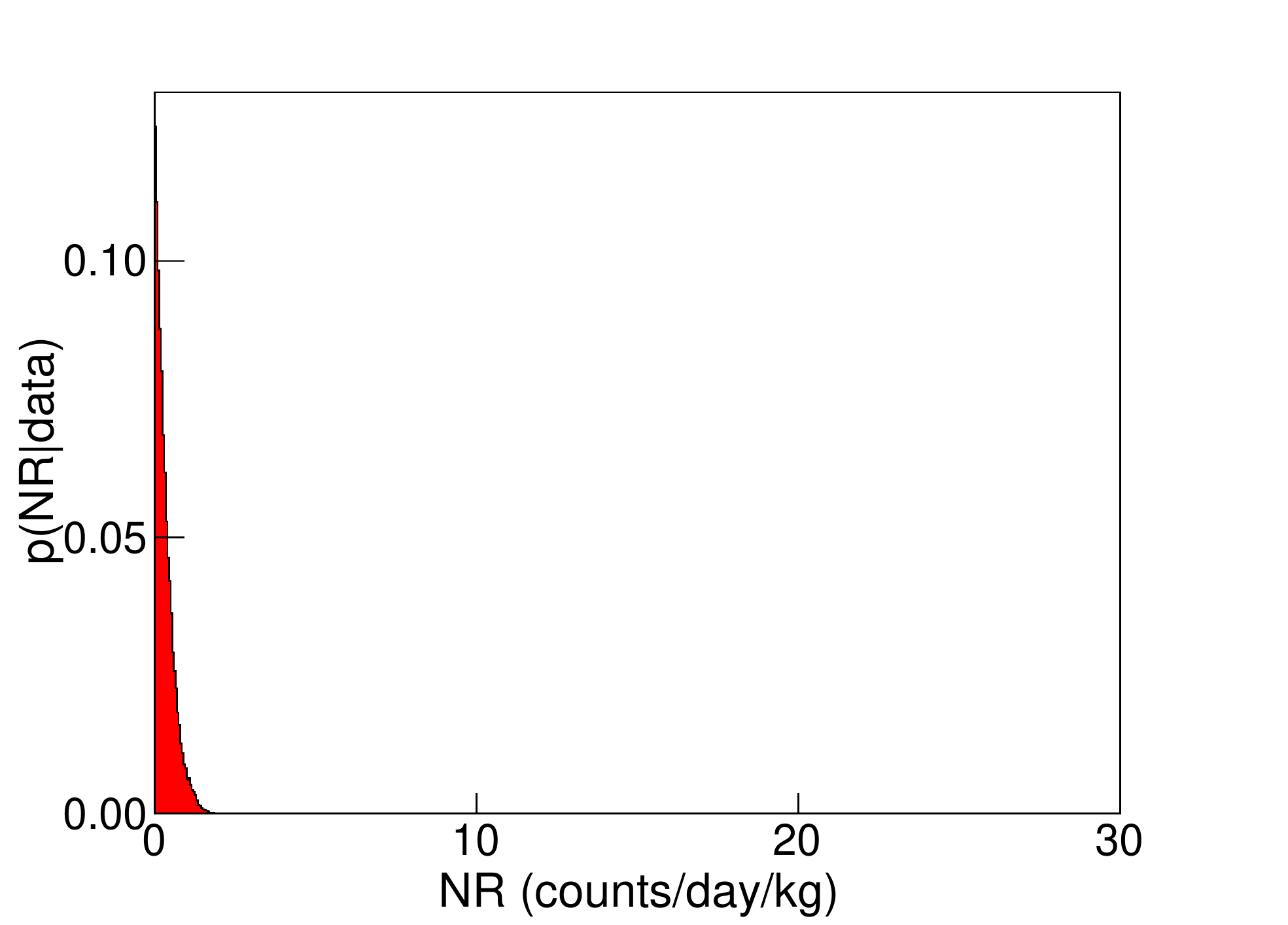} &
\includegraphics[width=0.4\textwidth]{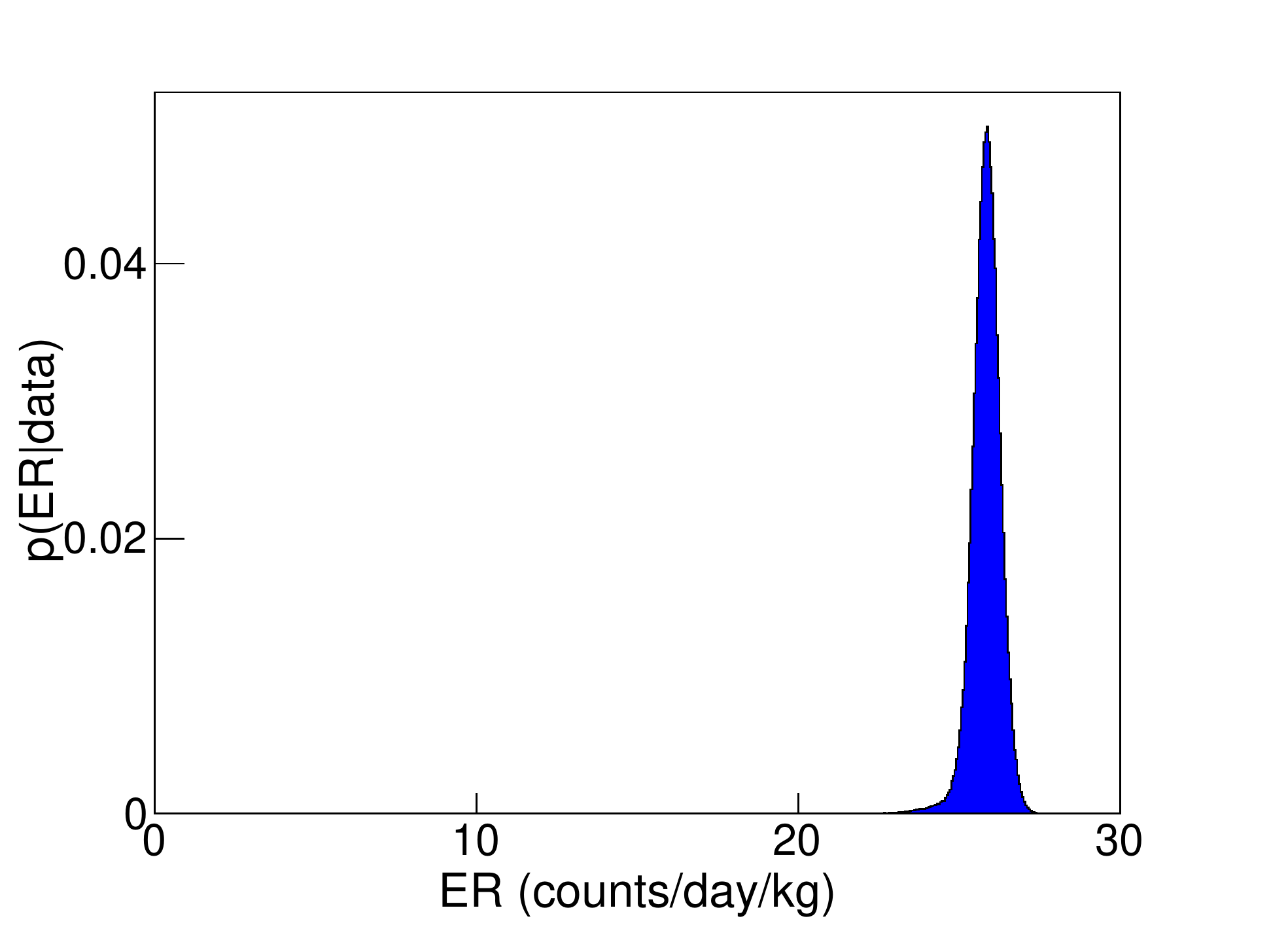} \\
(a) & (b) \\
\includegraphics[width=0.4\textwidth]{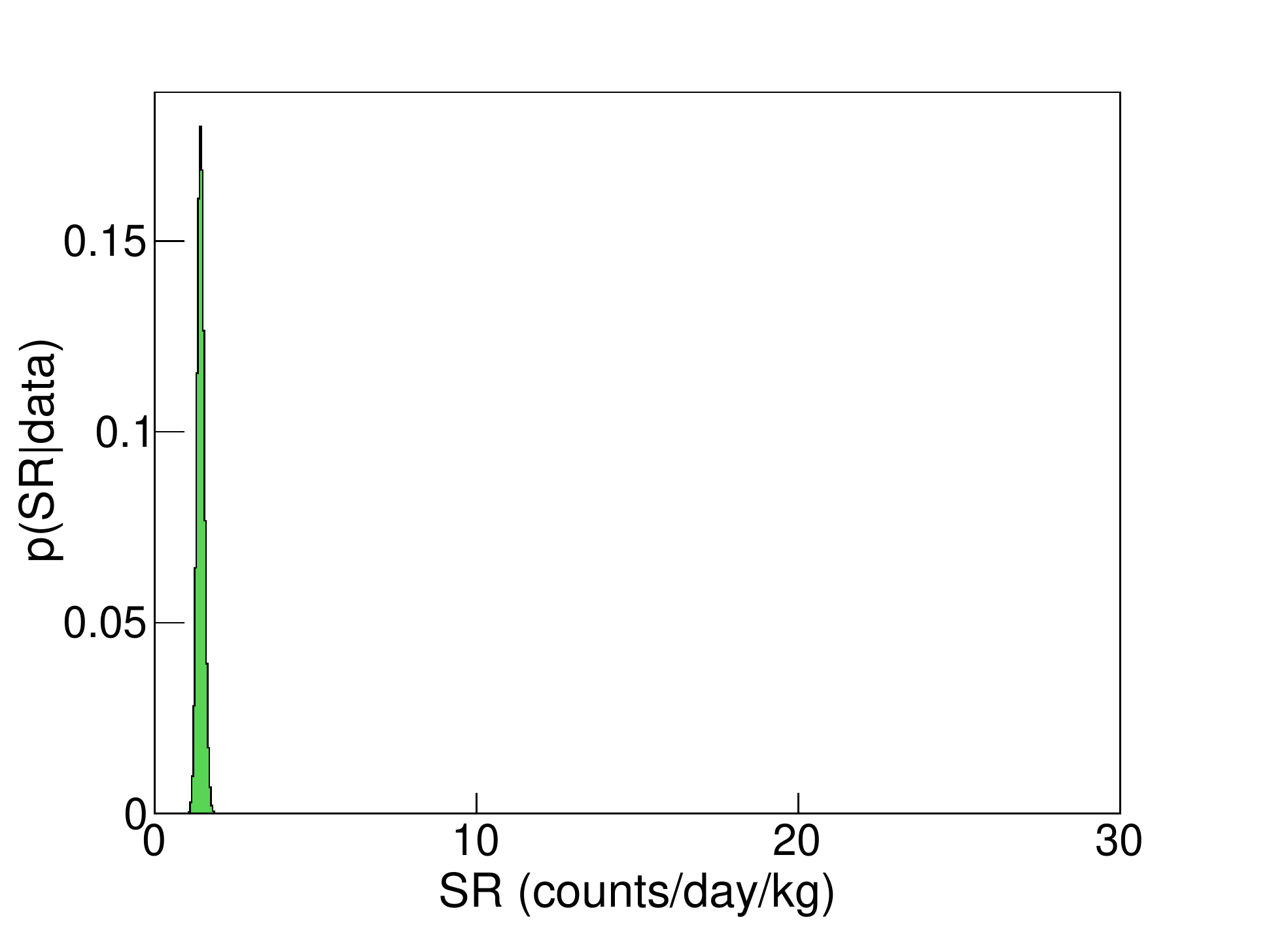} &
\includegraphics[width=0.4\textwidth]{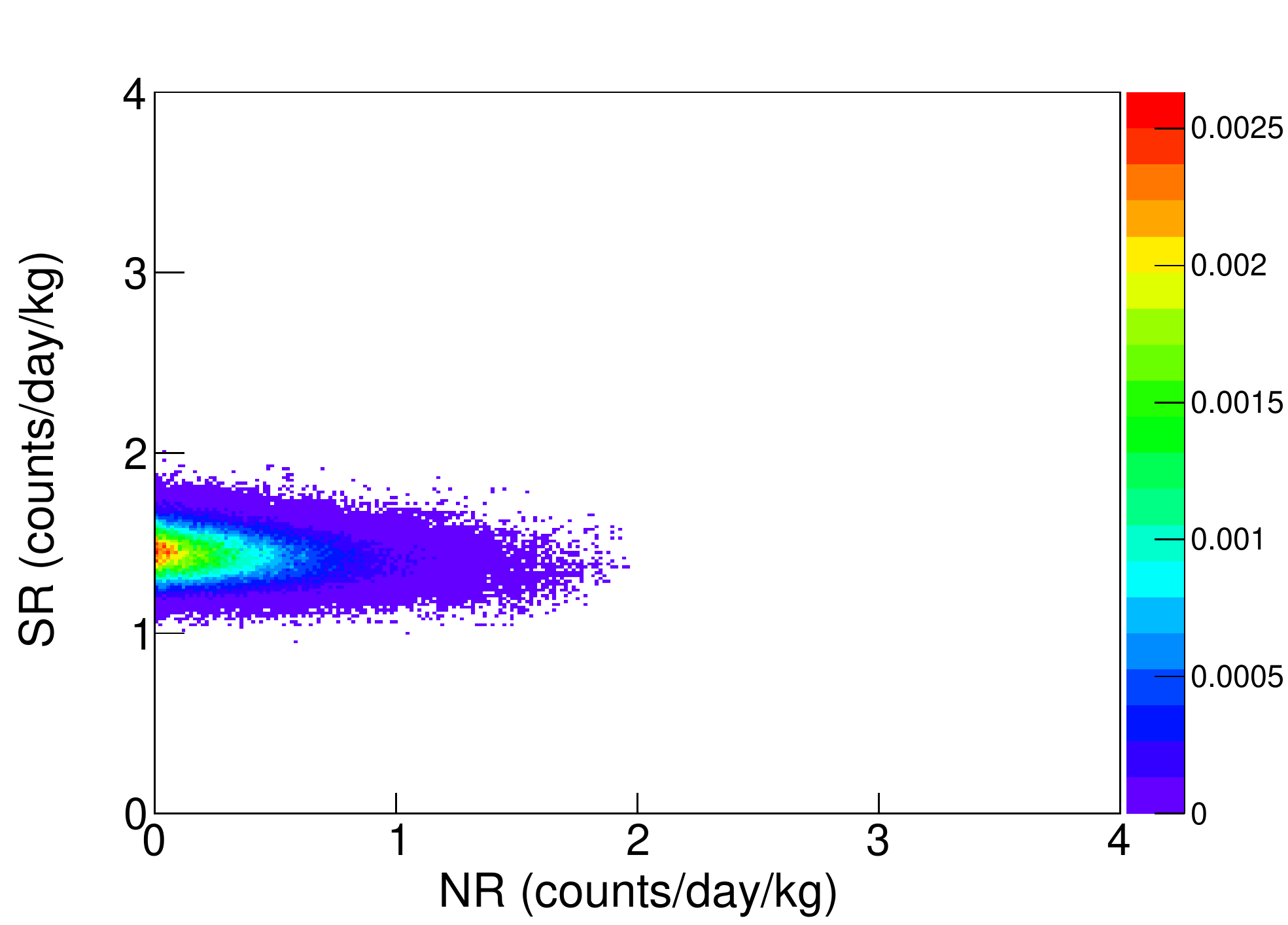} \\
(c) & (d) \\
\end{tabular}
\caption{\label{fig_res_fit}
The posterior PDFs for: (a) NR events; (b) ER events; and (c) SR events for 2$-$8~keV of NaI2 normalized to time~(day) and mass~(kg). (d) Two-dimensional distribution of posterior NR and SR PDFs, where no correlation is evident.
}
\end{figure}

\section{Results}

\begin{figure}[tbp]
\centering
\includegraphics[width=0.6\textwidth]{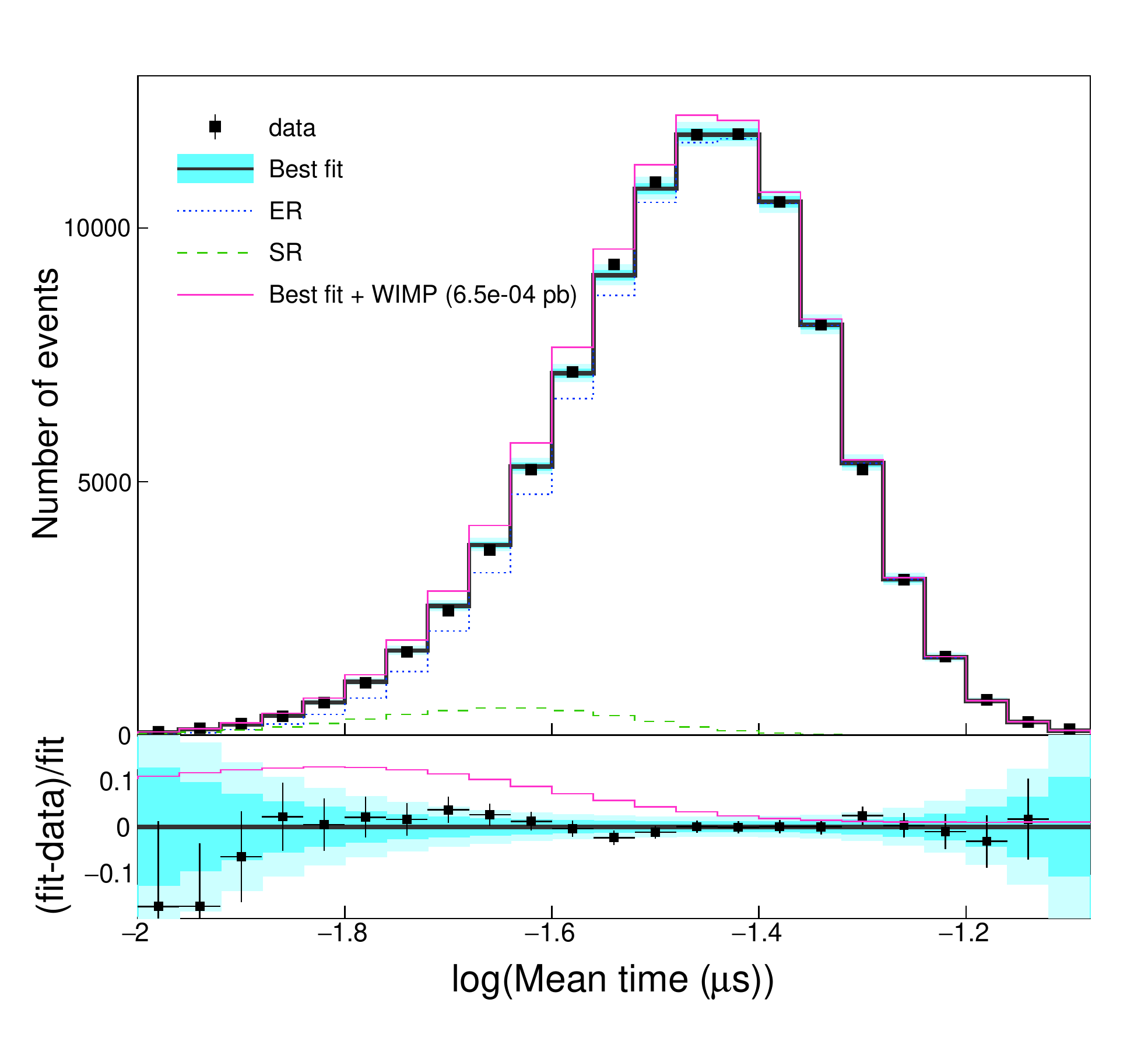}
\caption{\label{fig_res_mt}
The accumulated LMT spectrum for all energy bins in the ROI and the two crystals from the data~(square point) and best fit model~(solid line with shaded band corresponding to 1- and 2-$\sigma$ uncertainties).
The fitted results for ER events~(blue dotted line) and SR events~(green dashed line) are indicated.
The magenta thin solid line represents the expected signal excess above background for a 10~GeV/c$^2$ WIMP mass and $6.5\times10^{-4}$ pb cross section which corresponds to two times of 90\% limit value.
The lower panel shows the residuals between the data and the background model.
}
\end{figure}

Data fits are performed for each assumed WIMP mass value.
Figure~\ref{fig_res_fit}~(a) shows the posterior PDF of the NR rate in the case for a WIMP mass of 10~GeV/c$^2$ as an example.
The maximum value of the posterior PDF for the NR component is zero.
Figure~\ref{fig_res_fit}~(b) and (c) show those for the ER and SR backgrounds, which both have non-zero contributions.
The two-dimensional posterior PDFs for NR and SR are given in figure~\ref{fig_res_fit}~(d), where no evident correlation has been found.
The result of the maximum likelihood fit of the LMT spectrum for the 10 GeV/c$^2$ WIMP mass case is shown in figure~\ref{fig_res_mt}.
Here, the summed spectrum for the two crystals over the 2$-$8~keV ROI are shown together with the best-fit, non-NR signal result.
For comparison, the expected signal for a 10~GeV/c$^2$ WIMP mass and a spin-independent cross section of 6.5$\times10^{-4}$~pb which corresponds to two times of 90\% limit value added to the best fit is overlaid.

\begin{figure}[tbp]
\centering
\includegraphics[width=0.7\textwidth]{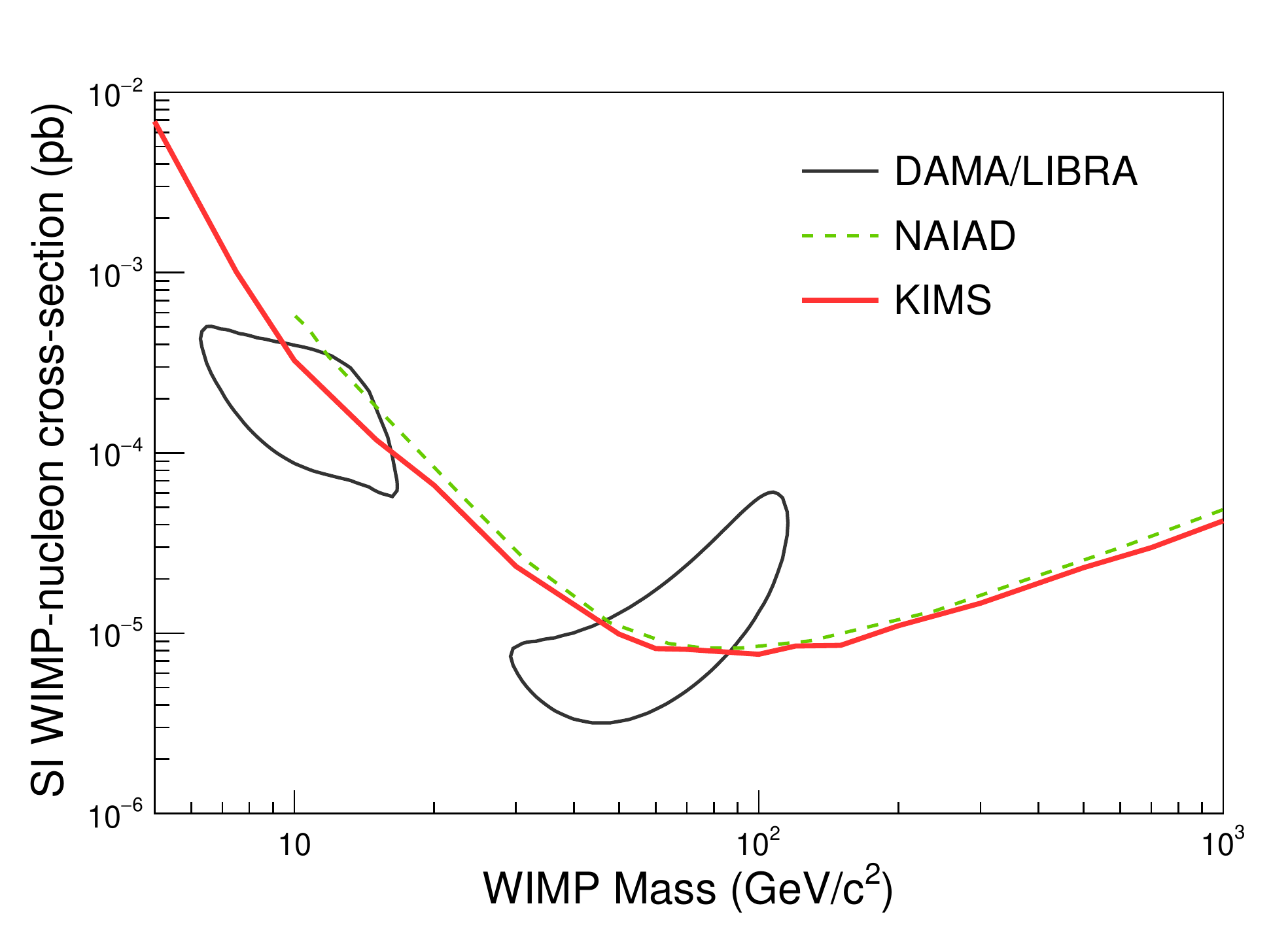}
\caption{\label{fig_limit}
Observed~(red solid line) 90\% CL exclusion limits on the WIMP-nucleon spin-independent cross sections.
The contour from WIMP-induced DAMA phase-1 3$\sigma$ allowed signal region~\cite{Savage:2008er} and the limit from the previous NaI(Tl) detector from the NAIAD experiment~\cite{Alner:2005kt} are shown.
}
\end{figure}

For all 17 WIMP mass values that are considered, the likelihood fits found no excess corresponding to NR signal events; all the posterior PDFs for the NR rates are consistent with zero.
We determine 90\% CL upper limits on spin-independent WIMP-nucleon cross sections, such that 90\% of the posterior densities fall below the limit, as shown in figure~\ref{fig_limit}.
The obtained limits are compared with the 3$\sigma$ contours of the allowed regions associated with the DAMA signal and the limits from the NAIAD experiment~\cite{Alner:2005kt} that used the NaI(Tl) crystals with 16388.5~kg$\cdot$days exposure using the PSD analysis.
Even though data exposure is only 20\% of the NAIAD experiment, we obtain overall slightly better limits by taking advantage of lower background and higher light yields that provided better PSD of NR events~\cite{Lee:2015iaa}.
The DAMA modulation signal was interpreted in WIMP parameter space by fitting the observed modulation amplitude to the expected spectrum from standard halo model  by assuming $v_{0}$ = 220 km/s, $\rho_{DM}$ = 0.3 GeV/$\text{cm}^{3}$, $v_{esc}$=650 km/s with parameters; WIMP mass and cross section.

\section{Conclusion}
We report limits on WIMP-nucleon cross sections using a data sample collected with two low-background NaI(Tl) crystals with a total exposure of 2967.4~kg$\cdot$days.
No excess of the NR events is observed, and 90\% CL upper limits on the WIMP-nucleon cross section are set.
The limit, 3.26$\times$10$^{-4}$ pb for a WIMP mass at 10~GeV/c$^2$, partially covers the DAMA 3$\sigma$ region of WIMP-sodium interaction and demonstrates the feasibility of using PSD techniques for the direct extraction of WIMP-nucleon scattering from ongoing NaI(Tl) experiments such as COSINE-100~\cite{cosine_set1} and ANAIS-112~\cite{Coarasa:2017aol}.

\acknowledgments
We thank the Korea Hydro and Nuclear Power (KHNP) Company for providing the underground laboratory space at Yangyang.
This research was supported by the Institute for Basic Science (Korea) under project code IBS-R016-A1.

%\paragraph{Note added.} This is also a good position for notes added after the paper has been written.


\bibliographystyle{JHEP}
\bibliography{dm}

%\begin{thebibliography}{99}

%\bibitem{a}
%Author, \emph{Title}, \emph{J. Abbrev.} {\bf vol} (year) pg.

%\bibitem{b}
%Author, \emph{Title},
%arxiv:1234.5678.

%\bibitem{c}
%Author, \emph{Title},
%Publisher (year).

% Please avoid comments such as "For a review'', "For some examples",
% "and references therein" or move them in the text. In general,
% please leave only references in the bibliography and move all
% accessory text in footnotes.

% Also, please have only one work for each \bibitem.

%\end{thebibliography}

\end{document}